\documentclass[a4paper,11pt]{article}
\usepackage{jheppub} 
\usepackage{amsmath}
\usepackage{tikz}
\usepackage{empheq}
\usepackage{mdframed}
\usepackage{esint}
\usepackage{bm}
\usepackage{simpler-wick}

\usetikzlibrary{positioning, shapes.geometric}
\usetikzlibrary{decorations.pathmorphing,arrows.meta}
\usetikzlibrary{calc, decorations.markings}
\usetikzlibrary{patterns}
\usetikzlibrary{patterns.meta}

\def\be{\begin{equation}}
\def\ee{\end{equation}}

\def\k{\textbf{k}}

\definecolor{shadecolor}{rgb}{0.92,0.92,0.92}

\begin{document}

\title{Cosmology, Decoherence and the Second Law}
	\author[a]{Sebasti\'an C\'espedes,}  
    \author[b]{Senarath de Alwis}  
    \author[c,d]{and Fernando Quevedo}  
	\affiliation[a]{Abdus Salam Centre for Theoretical Physics,\\ Imperial College London,
London, SW7 2AZ, UK}
    \affiliation[b]{Physics Department, University of Colorado, Boulder, CO 80309 USA}
    \affiliation[c]{New York University Abu Dhabi, PO Box 128199, Saadiyat Island, Abu Dhabi, UAE.}
    \affiliation[d]{DAMTP  University of Cambridge, Wilberforce Road,  Cambridge, CB3 0WA, UK}
\abstract{We consider quantum decoherence and entropy increase in  early universe cosmology.
  We first study decoherence in a  discrete bipartite quantum system for which  a single qubit  gets entangled with an environment and  the entropy increase is correlated with the 
decay of the  off-diagonal terms of the reduced density matrix. We compare this system with continuous systems  relevant for cosmology  for which there is a natural external intervention, corresponding to the time-dependent  separation between the sub-
and super-horizon inflationary fluctuations. We  find, in this case,  that the off-diagonal terms  of the density matrix, in a field basis, do not decay as sometimes assumed in cosmological set-ups. Nevertheless, following a recent treatment in terms of open Effective Field Theories (EFTs), we compute the entanglement entropy for a Gaussian state  and show that it  actually increases monotonically ($\dot S>0$) during the accelerated phases ($\ddot a>0$ with $a(t)$ the scale factor). We  generalise this result to include  non-Gaussian states and briefly discuss the relevance of computing the von Neumann entropy as compared to the thermodynamic entropy.
 }

\maketitle

\pagebreak


\section{Introduction}

Early-universe cosmology has long been considered the natural regime in which to address the origin of the second law of thermodynamics and the associated arrow of time (see, e.g., \cite{Penrose:1980ge,Wald:2005cb}). Naively, the observed increase of entropy today can be traced back to the assumption that the entropy of the early universe was much smaller. This leads to a fundamental question: why was the entropy so low at early times? Inflation itself does not provide an explanation but makes this question more interesting and challenging. Several proposals have been made to address this question but it is fair to say that it remains open (see for instance \cite{Carroll:2004pn,Bousso:2011aa}). 

In this note, rather than proposing a new general mechanism to solve this difficult problem, we focus on trying to make progress by computing the entanglement entropy in a standard cosmological set-up and consider its time evolution~\footnote{For other approaches to entanglement entropy in cosmology see for instance \cite{Wall:2010jtc,Maldacena:2012xp,Brahma:2020zpk, Boutivas:2023mfg,DuasoPueyo:2024usw}.}. It is well known that in a quantum mechanical system/environment set-up, an entanglement entropy can be computed by tracing out the contribution from the environment, breaking unitarity and allowing the possibility to have  entropy increase \citep{Zurek:2003zz}.

Cosmological horizons provide a natural way to separate a system from its environment: the sub-horizon modes form the ``system'' while the super-horizon modes act as the ``environment''. Here, following recent developments in open-system EFTs \citep{Calzetta:1995ys,Burgess:2006jn,Burgess:2015ajz,Boyanovsky:2015tba,Boyanovsky:2015xoa,Burgess:2022nwu,Colas:2022kfu,Salcedo:2024smn,Salcedo:2025ezu}, we compute the entanglement entropy of density-fluctuation modes and show that it grows monotonically as long as the universe undergoes accelerated expansion.

Let us make this point more explicit. 
The von Neumann entropy for a state described by a density matrix
$\rho$ is defined in analogy with the Boltzmann-Gibbs entropy in
classical physics,
\begin{equation}
S_{{\rm vN}}=-{\rm tr}\rho\ln\rho=-\sum_{i}p_{i}\ln p_{i}\label{eq:liouville}
\end{equation}
 where the second equality obtains in a basis in which $\rho$ is
diagonalized. If the state is pure ($\rho^2=\rho$), this entropy is zero. A  non-zero
entropy obtains by coarse graining, i.e. averaging over a bundle of microscopic variables, or  by considering a subsystem of a closed system and tracing over all observables in its complement (the ``environment"). We will be discussing the latter situation, in other words the physics of an open system coupled to an environment. 
The entropy of the open system is then defined in terms of,
\begin{equation}
\rho_{{\rm red}}\left(\alpha\right)={\rm tr}_{\beta}\rho\left(\alpha,\beta\right)\label{eq:rhored}
\end{equation}
 where the set $\alpha$ are the observables in the subsystem
 and $\beta$ are those of the environment. 
Since in general $\rho^{2}\ne\rho$, then the von Neumann entropy is different from zero, $S_{{\rm vN}}\ne0$.


A pure state or even a statistical (time independent) mixture of pure states evolves
unitarily. i.e. satisfies (working in the Schr\"odinger picture) the
quantum version of Liouville's equation 
\begin{equation}
i\hbar\frac{d}{dt}\rho=\left[H,\rho\right].\label{eq:quantLiouville}
\end{equation}
 This implies that the corresponding entropy (even if non-zero) remains
constant in time. Thus even though \eqref{eq:liouville} has non-zero
entropy there is no second law in the form of an inequality since
$dS_{{\rm vN}}/dt=0$. However, this is not necessarily the case for the entanglement entropy
$$S_{{\rm EE}}=-{\rm tr}\rho_{\rm red}\ln\rho_{\rm red}$$ defined using the reduced densty matrix
\eqref{eq:rhored}, since in general, it does not  evolve unitarily and therefore equation \eqref{eq:quantLiouville} should be modified by adding a second term in the RHS as in a master equation for which probability distributions change with time.
This distinction is crucial for the rest of our discussion. 

In our cosmological application, where we assume that the fluctuations for the CMB modes were produced during inflation, the  division between system and environment can be naturally thought of as provided by the comoving Hubble scale: modes with $k \leq \Lambda(t)$ are ``system'' modes, while those with $k > \Lambda(t)$ are ``environment'' modes, with
$\Lambda(t) = a(t) H$~\cite{Lombardo:2005iz,Campo:2008ju,Campo:2008ij,Burgess:2014eoa,Burgess:2020tbq,Brahma:2020zpk,Burgess:2022nwu}. 
We show that the entanglement entropy of the system modes grows monotonically during accelerated expansion ($\ddot{a} > 0$) for both Gaussian and weakly non-Gaussian states.\footnote{These results can be generalised to other theories provided that $\ddot a>0$, see \cite{Colas:2024xjy}.}

We first illustrate entropy increase in a discrete bipartite quantum model where a single qubit interacts with an environment, demonstrating how entropy growth correlates with the decay of off-diagonal elements in $\rho_{\rm red}$. We then extend the analysis to cosmology, highlighting the different role of the off-diagonal density-matrix elements in the inflationary context. We close with a discussion of how entanglement entropy compares to standard thermodynamic entropy, especially after the end of acceleration. An appendix outlines a general path integral QFT approach for computing purity and entropy in open systems.

In concluding this introduction we emphasize that we are working in a field theory with a UV cutoff, so that all physical momentum modes $k/a \ll M_P$. Furthermore our 
calculations are done effectively in a finite volume universe which we may send to infinity whenever that is required and well defined. Thus issues that plague attempts to formulate cutoff  free QFT (such as the non-existence of the interaction picture) are not relevant since we are of the opinion that somewhere close to the Planck scale QFT should be replaced, perhaps by string theory.

\section{Decoherence and entropy increase}
\label{sec:decoherence}

In quantum mechanics, decoherence refers to the loss of quantum coherence of a system due to the interaction with the environment (for a review see for instance \citep{Zurek:2003zz}).  In quantum field theory, the phenomenon of decoherence is closely related to what is now called open effective field theories (EFTs). Open EFTs differ from standard EFTs in the sense that the states that are integrated out are not necessarily of higher energies but  depend on the separation between the system and the environment (see \citep{Burgess:2020tbq} for a review).

We will address below how decoherence can be closely related to the entropy increase in quantum mechanical systems.

\subsection{Decoherence in discrete  bipartite systems}

Consider a system $S$ with states $|s_{i}\rangle,\,i=1,\ldots n$ interacting
with an environment ${\cal E}$ with states $|e_{\alpha}\rangle,\,\alpha=1,\ldots,N$,
(with $N\gg n$). The Hamiltonian is assumed to be of the form

\begin{equation}
\hat{H}=\hat{H}_{S}\otimes\hat{I}_{{\cal E}}+\hat{I}_{S}\otimes\hat{H}_{{\cal E}}+ \hat{H}_{S{\cal E}}.
\label{eq:H}
\end{equation}
Here the first(second) term acts on the system(environment) and the
third term is the interaction that can be written as $\hat{H}_{S{\cal E}}
=\sum_{m}g_{m}\hat{h}_{m}\otimes\hat{f}_{m} $ where the first operator  in the product, $\hat h_m$,
acts only on the system and the second, $\hat{f}_m$, on the environment, with $g_m$ the coupling constants. At time
$t=0$ the system states are assumed to be uncorrelated with the environment,
i.e.
\[
|\psi\left(0\right)\rangle=\sum c_{i}(0)|s_{i}\rangle |e_{0}\rangle.
\]

Let us now specialize to an example discussed by Zurek \citep{Zurek:2003zz}
for environment induced superselection. The model consists of a single
qubit system (so $n=2$)  interacting with  an environment consisting
of $N$ qubits. In this model we ignore the internal dynamics of each
subsector so the Hamiltonian is just the interaction Hamiltonian. 
\begin{equation}
\hat H=\hat H_{S{\cal E}}=\sigma_{z}^{S}\otimes\sum_{i}\left(g_{i}\sigma_{z}^{{\cal E}i}\otimes\prod_{i'\ne i}I^{{\cal E}i'}\right).\label{eq:HIzurek}
\end{equation}
With $\sigma_z$'s,  the Pauli matrices. The initial state is taken to be (with $|\pm\rangle$ the eigenstates of
$\sigma_{z}$)
\begin{align}
|\psi\left(0\right)\rangle & =\left(c_{1}|\, +\, \rangle +c_{2}|-\rangle \right)|e_{0}\rangle ,\qquad\quad \,  |c_{1}|^{2}+\,|c_{2}|^{2}=1\label{eq:psi0}\\
|e_{0}\rangle & =\prod_{k=1}^{N}\left(\alpha_{k}|+\rangle_{k}+\beta_{k}|-\rangle_{k}\right),\qquad |\alpha_{k}|^{2}+|\beta_{k}|^{2}=1.\label{eq:e0}
\end{align}

Carrying out the unitary evolution under this Hamiltonian 
\begin{equation}
|\psi\left(t\right)\rangle=c_{1}|+\rangle |e_{0}(t)\rangle+c_{2}|-\rangle |e_{1}(t)\rangle,
\label{eq:psit}
\end{equation}
 where 
\begin{equation}
|e_{0}\left(t\right)\rangle=\prod_{k=1}^{N}\left(\alpha_{k}e^{ig_{k}t}|+\rangle_{k}+\beta_{k}e^{-ig_{k}t}|-\rangle_{k}\right)=|e_{1}\left(-t\right)\rangle .\label{eq:eoft}
\end{equation}
The reduced density matrix  for the system $\rho_{_S} $ is then obtained after tracing over the environment
\begin{align}
 \rho_{_S}= {\rm tr}_{_{\cal E}}\rho= & |c_{1}|^{2}|+\rangle \langle+|\, \, +\, \, |c_{2}|^{2}|-\rangle \langle-|\nonumber \\
 & +c_{1}c_{2}^{*}\, r(t)|+\rangle\langle -|\,\, +\,\, c_{1}^{*}c_{2}\, r^{*}(t)|-\rangle\langle+|. \label{eq:rhoS}
\end{align}

Here the overlap between the selected (by the system) of the environment
states is (see also \citep{PhysRevA.72.052113}):
\begin{align}
  r(t) & \equiv \langle e_{0}\left(t\right)|e_{1}(t)\rangle =\prod_{k=1}^{N}\left[|\alpha_{k}|^{2}\exp\left(i2g_{k}t\right)+|\beta_{k}|^{2}\exp\left(-i2g_{k}t\right)\right] =\sum_{n=0}^{2^{N}-1}|c_{n}|^{2}e^{-iB_{n}t}\nonumber \\
 & =\int_{-\infty}^{\infty}e^{-iBt}\eta\left(B\right)dB,\qquad {\rm with}\qquad \eta(B)=\sum_{n=0}^{2^{N}-1}|c_{n}|^{2}\delta(B-B_{n}).\label{eq:e0e1}
\end{align}
Here we have defined $B_{n}=\sum_{k=1}^{N}\left(-1\right)^{n_{k}}2g_{k},$ with $n_{k}=(1-(-1)^n)/2$. Also $ \sum_{n=0}^{2^{N}-1}|c_{n}|^{2}=1$.

The late time suppression of the overlap $r(t)$ by suitable choices of $\eta(B)$ has been shown in various
models. See for instance \citep{PhysRevA.72.052113}. Below we will show how this leads to the second law.


We now  compute the eigenvalues of the reduced density matrix  $\rho_{_S}(t)$ given in \eqref{eq:rhoS}, by solving the equation  $\det({\rho}_{_S}(t)-\lambda\hat{I})=0$,
\begin{equation}
\lambda=\lambda_{\pm}=\frac{1}{2}\left\{ 1\pm\left(1-4|c_{1}|^{2}|c_{2}|^{2}\left(1-|r(t)|^{2}\right)\right)^{1/2}\right\} \label{eq:lambdaplusmiinus}
\end{equation}
Note that  at $t=0$ we get $r(t=0)=1$, $\lambda_{+}=1$ and $\lambda_{-}=0$
as should be the case since initially the system is in a pure state.
Also $\frac{1}{2}<\lambda_{+}\le1$ and $0\le\lambda_{-}\le\frac{1}{2}$. But as remarked above, for weak enough couplings $g_k$, $r(t)$ decreases with time although for very long times the periodicity of the exponential functions may drive $r(t)$ back towards $r=1$ as expected from the quantum Poincar\'e recurrence theorem..
\subsection{Entropy increase}
Computing the entropy we have (note that $\lambda_{+}+\lambda_{-}=1)$
\begin{equation}
S_{S}=-{\rm tr}\rho_{S}\ln\rho_{S}=-\lambda_{+}\ln\lambda_{+}-\left(1-\lambda_{+}\right)\ln\left(1-\lambda_{+}\right)\geq 0.\label{eq:Ssystem}
\end{equation}
 Using \eqref{eq:lambdaplusmiinus} we then have 
\[
\frac{dS_{S}}{dt}=\ln\left(\lambda_{+}^{-1}-1\right)\frac{2|r|\vert\dot{r}\vert|c_{1}|^{2}|c_{2}|^{2}}{\left(1-4|c_{1}|^{2}|c_{2}|^{2}(1-|r|^{2})\right)^{1/2}}.
\]
Note that the first factor is negative since $\lambda_{+}^{-1}<2$.

Therefore we can see that
\begin{equation}
\frac{dS_{S}}{dt}>0 \iff \frac{d|r|}{dt}<0
\end{equation}
So a strict ``second law'' $\frac{dS_{S}}{dt}>0$ is obtained as long as  $\frac{d|r|}{dt}<0$.
In other words this ``second law'' is equivalent to  the decay of the overlap
between the two environment states which in turn is the manifestation of decoherence. Crucially this implies that the density matrix ends up being diagonal. Therefore illustrating the connection between decoherence and the second law.


However as noted for instance in \citep{PhysRevA.72.052113}, as long as $N$ is finite the models which
led to the decay of $r(t)$ 
cannot be valid
for all time $t.$ The reason is that for finite $N$ the expression
for $r(t)$ \eqref{eq:e0e1} is a finite sum of periodic functions
and hence is an almost periodic function and therefore can for late
times come back (or almost come back) to its initial value an infinite
number of times. This is related to the quantum version of a Poincar\'e
recurrence. 
Suffice it to
say here that for large enough $N$ this recurrence
(and hence recoherence) will take place at a time comparable to the
age of the universe, so that for time scales much shorter than that
and for large enough environments, decoherence and hence the related
increase in entropy will be operative.

However this ``second law'' is clearly different from the thermodynamic
second law. The latter applies
to a closed system and not to an open one such as the subject of decoherence
theory. Furthermore the two subsystems (system and environment) have
the same von Neumann entropy if the total system is pure. This means
that both entropies must increase (or decrease if there is recoherence).

This is very different from the behaviour of a thermal system that
is moving towards equilibrium. Here, if initially there is a constraint
dividing the (closed) system into subsystems - upon the removal of
the constraint the entropy of the high entropy state decreases and
the other increases (think of the cooling of a coffee cup or the removal
of a partition of a insulated box containing air on one side of a
partition but not the other), such that the total entropy increases.

Finally, let us emphasize that the decoherence argument does not, by itself, introduce an ``arrow of time''. At the initial moment (say, $t = 0$), the system is assumed to be in a pure state, unentangled with the environment. It is the Hamiltonian evolution that generates entanglement and thereby leads to an increase in the entropy of both the system and the environment. However, Hamiltonian evolution is time-reversal symmetric. Consequently, the same mechanism also applies for $t < 0$.\footnote{Note that the models for the environmental correlations are likewise time-reversal symmetric.
} The system would therefore become entangled and gain entropy as one moves backward in time as well. This apparent paradox can only be avoided if one postulates a special initial state (such as an uncorrelated initial state assumed in decoherence arguments) defined at the beginning of time.

\section{Entropy increase in quasi de Sitter cosmology}
\label{sec:entropy_inflation}

\subsection{Soft vs hard modes and the diffusion equation}

As discussed in the previous sections, when the separation between the system and its environment is fixed, the entropy obtained by tracing over the environmental degrees of freedom remains constant or increases if the overlap between certain environment states $r(t)$ decays with time. However, in cosmology - as well as in black hole evaporation- this separation is naturally time-dependent. To account for this, we divide the CMB modes into high-momentum modes $k > \Lambda(t)$ and low-momentum modes $k \leq \Lambda(t)$, where the cut-off $\Lambda(t)$ evolves due to cosmological expansion. A natural choice for the cut-off is $\Lambda = \epsilon a H$, where a is the scale factor, $H = \dot{a}/a$ is the Hubble parameter, and $\epsilon$ is a free parameter (note that for $\epsilon = 1$, we recover the usual separation between long and short modes).

The entropy $S$ associated with the soft modes will thus satisfy
a time dependence of the form
\begin{equation}
\frac{dS}{dt}=\left.\frac{\partial S}{\partial t}\right\vert_{\Lambda}+\dot{\Lambda}\frac{\partial S}{\partial\Lambda}.\label{eq:dSbydt}
\end{equation}
The first term corresponds to the time evolution for a fixed separation
between the environment (hard modes) and the system (soft modes) and
which, as we argued in the previous sections, is positive or zero,
and the second term comes from the time dependence of the cut-off.
As argued in \citep{Gorbenko:2019rza} while the first term corresponds
to the dynamical evolution of the system the second may be identified
with the diffusion term in the Fokker-Planck equation.

\subsubsection*{Diagonal density matrix}

Let us define the quantum state of the CMB fluctuations as ${\rho}=|\Psi\rangle\langle\Psi|$.
The corresponding reduced density matrix of the superhorizon modes
is then given by\begin{equation}
{\rho}_{\Lambda}=\int\prod_{k>\Lambda}d\phi_{k}\langle \phi_{k} |\Psi\rangle\langle\Psi| \phi_{k}\rangle\label{eq:rhoLambda}
\end{equation}
In the absence of interaction this reduced density matrix is still
a pure state. To generate entanglement between the UV and the IR modes
we assume that there is a weak coupling between them. For the sake
clarity we assume that the spatial volume is finite so that the momentum
modes are discrete.
For the case of a single field $\phi$ and a probability density for soft modes $P_{\Lambda}\left(\phi\right)$ with
(below we take the momentum modes to be dimensionless and discrete\footnote{We take space to be compactified on a torus with volume $V_{\rm space}=L^3 $ so that ${\bf k}=2\pi{\bf n}/L$. At the end of the calculation we will take the large volume limit.}
 to be consistent
with the formalism of \citep{Cespedes:2023aal})
\begin{equation}
\int\prod_{k\le\Lambda}d\phi_{k}P_{\Lambda}\left(\phi\right)=1,\qquad \,P_{\Lambda}\left(\phi\right)=\int\prod_{k>\Lambda}d\phi_{k}P\left(\phi\right)\label{eq:PLambda}
\end{equation}
The quantum (von Neumann) entropy of the soft modes is given by 

\begin{align}
S_{\le\Lambda} & =-\int\prod_{k\le\Lambda}d\phi_{k}\langle \phi_{k} |{\rho}_{\Lambda}\ln{\rho}_{\Lambda}| \phi_{k} \rangle\nonumber \\
 & =-\int\prod_{k\le\Lambda}d\phi_{k}\langle\phi_{k} |{\rho}_{\Lambda}\int\prod_{k<\Lambda}d\phi'_k| \phi'_{k} \rangle\langle \phi'_{k} |\ln{\rho}_{\Lambda}| \phi_{k} \rangle\nonumber \\
 & =-\int\prod_{k\le\Lambda}d\phi_{k}\langle\ \phi_{k} |{\rho}_{\Lambda}| \phi{}_{k} \rangle\ln\left(\langle \phi_{k} |{\rho}_{\Lambda}| \phi_{k} \rangle\right)+\ldots\nonumber \\
 & =-\int\prod_{k\le\Lambda}d\phi_{k}P_{\Lambda}\left(\phi\right)\ln P_{\Lambda}\left(\phi\right)+{\rm off\,diagonal\,terms.}\label{eq:Stodiagonal}
\end{align}

As discussed in Sec.\eqref{sec:decoherence} and also commonly argued  in the decoherence literature \cite{Zurek:2003zz}, the off-diagonal terms
are estimated to be suppressed compared to the diagonal terms. 
If this is the case here, we may approximate the
exact entropy of the state of soft modes by a semi-classical entropy
(similar to the classical Boltzmann-Gibbs entropy) by 
\[
S_{{\rm sc}}=-\int d\phi_{l}P_{\Lambda}\left(\phi_l\right)\ln P_{\Lambda}\left(\phi_l\right).
\]
Then we have (since $\int d\phi_{l}\frac{d}{d\ln\Lambda}P_{\Lambda}\left(\phi_l\right)$=0) that
\[
\frac{\partial S}{\partial\ln\Lambda}=-\int d\phi_{l}\frac{\partial P_{\Lambda}\left(\phi_l\right)}{\partial\ln\Lambda}\log P_{\Lambda}\left(\phi_l\right).
\]
The $\Lambda$ derivative of $P_{\Lambda}$ is related to the diffusion
term in the Fokker-Planck equation \citep{Gorbenko:2019rza,Cespedes:2023aal},
and we have taken the continuum limit (see for example equation 5.21 of \citep{Cespedes:2023aal}),
\begin{equation}
\frac{\partial P_{\Lambda}\left(\phi\right)}{\partial\ln\Lambda}=V_{{\rm space}}\frac{H^{2}}{8\pi^{2}}\int\frac{d^{3}k}{(2\pi)^{3}}\frac{1}{k^{3}}\frac{d\Omega_{\Lambda}}{d\ln\Lambda}\frac{\partial^{2}P_{\Lambda}}{\partial\phi_{k}\partial\phi_{k}^{*}}\label{eq:dPdL}
\end{equation}
(with $\Omega_{\Lambda}$ being a smooth window function that cuts
off the modes $k>\Lambda$) and hence\footnote{In a systematic derivation of the Fokker-Planck equation in this context
from quantum field theory in de Sitter space \citep{Cohen:2021fzf}
there are higher order derivative terms (associated with non-Gaussianities)
which are however suppressed compared to the leading term.}\footnote{Note that $\phi_{k}$ are the Fourier components of the fluctuations
of the inflaton about its classical background value which determines
$H$.}
\begin{align}
\frac{\partial S}{\partial\ln\Lambda} & =-V_{{\rm space}}\frac{H^{2}}{8\pi^{2}}\int\prod_{k'\le\Lambda}d\phi_{k'}\int\frac{d^{3}k}{(2\pi)^{3}}\frac{1}{k^{3}}\frac{d\Omega_{\Lambda}}{d\ln\Lambda}\frac{\partial^{2}P_{\Lambda}}{\partial\phi_{k}\partial\phi_{k}^{*}}\ln P_{\Lambda}\nonumber \\
 & =V_{{\rm space}}\frac{H^{2}}{8\pi^{2}}\int\prod_{k'\le\Lambda}d\phi_{k'}\int\frac{d^{3}k}{(2\pi)^{3}}\frac{1}{k^{3}}\frac{d\Omega_{\Lambda}}{d\ln\Lambda}\left|\frac{\partial P_{\Lambda}}{\partial\phi_{k}}\right|^{2}P_{\Lambda}^{-1}\nonumber \\
 & \simeq V_{{\rm space}}\frac{H^{2}}{16\pi^{4}\Lambda^{2}}\int\prod_{k\le\Lambda}d\phi_{k}\left|\left.\frac{\partial P_{\Lambda}}{\partial\phi_{k}}\right\vert_{k=\Lambda}\right|^{2}P_{\Lambda}^{-1}>0.\label{eq:dSbydL}
\end{align}

In the penultimate step we replaced the smooth window function by
a step function. Hence the second term in \eqref{eq:dSbydt} is positive
(or negative) if $\dot{\Lambda}>0$ (or $<0$). Given that 
the first term in \eqref{eq:dSbydt}, which is essentially just the classical entropy formula,  is expected to be non-negative\footnote{This depends on the validity of master equation arguments for this quantity as for example in standard discussions of classical statistical mechanics.}  this establishes an
increase of this semi-classical entropy if the universe is accelerating!

\begin{equation}
\frac{\dot{\Lambda}}{\Lambda}=\frac{1}{H}\left(H^{2}+\dot{H}\right)=\frac{\ddot{a}}{aH}\qquad {\rm then}\qquad \ddot a \geq 0 \implies \dot S \geq 0
\end{equation}.

As a consequence of this we see that the entropy associated with the
scalar fluctuations will rise during the inflationary phases - both
primordial and late time! However during the FRW (radiation and matter
dominated phases, i.e. when the universe is decelerating the sign of  $dS/dt$ will depend on the relative magnitudes of the two terms on the RHS of eqn. \eqref{eq:dSbydt}
. We will address these issues in a later section.

Unfortunately the above argument based on neglecting the off-diagonal terms (as happened in the discrete model above),  fails for continuum systems as it is not true that the off-diagonal components of the density matrix can be neglected, but as we will demonstrate assuming a Gaussian
state for the system, the story is more subtle. So far, we can only claim  that the  computation above shows that the diagonal part of the density matrix respects a second law. 

\subsection{Purity and Entropy of a Gaussian state }

In this section we construct the (Gaussian) reduced density matrix
following closely the discussion in \citep{Burgess:2022nwu}. Let
us take the following ansatz for the reduced density matrix defined
in \eqref{eq:rhoLambda},
\[
\langle \phi|{\rho}^{\Lambda}|\phi'\rangle=\prod_{{\bf k},k\lesssim\Lambda}\langle \phi_{{\bf k}}|{\rho}_{{\rm red}}|\phi'_{{\bf k}}\rangle,
\]
with $\phi_{k}$ being the Fourier components $\phi_{k}=\int\frac{d^{3}x}{\left(2\pi\right)^{3/2}}\phi\left(x\right)e^{-i{\bf k}\cdot{\bf x}}$ of
the field (the reality of $\phi$ implies $\phi_{-{\bf k}}=\phi_{{\bf k}}^{*}$).
We've also assumed above that the IR modes are uncorrelated initially.
Also following the literature (such as \citep{Burgess:2014eoa,Burgess:2022nwu}
the latter of which we follow closely in this discussion), we take the dominant interaction to be linear in the IR field. Under this assumption, each momentum mode evolves independently. Also let us define the real Fourier
component fields $\phi_{k}\equiv\left(\phi_{k}^{R}+i\phi_{k}^{I}\right)$,
and the matrix,

$$\boldsymbol{\Sigma}={\rm tr}{\rho}\begin{bmatrix}\phi^{2} & \left\{ \phi,\pi\right\} \\
\left\{ \phi,\pi\right\}  & \pi^{2}
\end{bmatrix}$$
 where $\phi,\pi$ stand for $\hat{\phi}_{k}^{R,I},\hat{\pi}_{k}^{R,I}.$  In the context of cosmology the  density matrix may be defined on the Schwinger-Keldysh contour where the time contour is doubled with the upper part representing time ordered evolution and the lower part anti time ordered evolution (See for example \cite{Kamenev_2011}). The time evolution of the field is given by the quantum state of the field,
 \begin{align}
 \vert \Psi(t_0)\rangle= U(t_0,-\infty)\vert\Omega\rangle
 \end{align} 
 where we have assumed that the fields begin  the adiabatic vacuum $\vert\Omega\rangle$ at the asymptotic past $t\to-\infty$ and $U(t_0,-\infty)$ is the time evolution operator derived from the Hamiltonian of the system. The wavefunction of the universe is defined by taking the initial state to be Bunch-Davies(BD),
 \begin{align}
 \Psi(\phi,t_0)\equiv\langle\phi\vert\Psi(t_0)\rangle=\int_{\rm BD}^{\varphi(t_0)=\phi}d[\varphi]e^{iS[\varphi;t_0]}
 \label{wvfn_def}
\end{align} 
Finally we can   define the density matrix as,
 \begin{align}
 \rho[\phi_+,\phi_-;t_0]\equiv \langle\phi_+\vert\rho(t_0)\vert\phi_-\rangle=\Psi[\phi^a;t_0]\Psi^*[\phi_b;t_0]
 \end{align} where the subindex $\pm$ specifies in which part of the contour the field is defined. The reduced density matrix between field eigenstates is obtained by integrating UV degrees of freedom. When the coupling between the short and long modes is linear we obtain at leading order a Gaussian density matrix  \footnote{A detailed derivation of how a Gaussian state emerges through coarse-graining of short-wavelength modes, using the path integral formalism, is presented in Appendix \ref{Appendix:CoarseGraining}.}, 
 
 \begin{equation}
\langle \phi^+_{kR}|{\rho}_{{\rm red}}|\phi^-_{kR}\rangle =\sqrt{\frac{1}{2\pi\Sigma_{k11}^{R}}}\exp\left(-\frac{a_{k}}{2}\left(\phi_{kR}^+\right)^{2}-\frac{a_{k}^{*}}{2}\left(\phi^-_{kR}\right)^{2}+c_{k}\phi^-_{kR}\phi^{+}_{kR}\right)\times\left(R\rightarrow I\right)\label{eq:Gaussian}
\end{equation}

Where the time dependent coefficients in the exponent are given by
(dropping the suffix $k$),
\begin{equation}
{\mathrm{Re}}\,    a=\frac{1}{\Sigma_{11}}\left[\det\Sigma+\frac{1}{4}\right],\qquad {\mathrm{Im}}\,     a=-\frac{\Sigma_{12}}{\Sigma_{11}},\qquad c=c^{*}=\frac{1}{\Sigma_{11}}\left[\det\Sigma-\frac{1}{4}\right].\label{eq:coeffs}
\end{equation}
We will now drop the superscripts $R,I$ 
until further notice and focus on the one of the factors in \eqref{eq:Gaussian}.

The purity of the state is easily computed and gives
\begin{equation}
\gamma_{k}:={\rm tr}\left[{\rho}_{k}^{2}\right]=\frac{1}{2\sqrt{\det\Sigma_{k}}}.\label{eq:purity}
\end{equation}
 Note that for a pure state $c=0$, i.e. $\det\Sigma=\frac{1}{4},$   so
the purity is unity (since $\rho^{2}=\rho)$. Purity less than unity arises only after tracing over some sector of the theory. For a Gaussian state
such as \eqref{eq:Gaussian} the quantum (i.e. von Neumann) entropy
can also be calculated (with somewhat more effort \citep{Serafini:2003ke})
and gives 
\begin{equation}
S=\frac{1-\gamma}{2\gamma}\ln\left(\frac{1+\gamma}{1-\gamma}\right)-\ln\left(\frac{2\gamma}{1+\gamma}\right),\qquad \gamma:=\frac{1}{2\sqrt{\det\Sigma}}.\label{eq:GaussianEntropy}
\end{equation}
As expected for a pure state i.e. $\det\Sigma=\frac{1}{4}$, $S=0.$
As we will see \citep{Burgess:2022nwu}, for late times $\eta\rightarrow-0$,
$\det\Sigma\rightarrow\infty$, so the purity goes to zero and the
entropy $S\rightarrow\infty.$ However as pointed out in \citep{Burgess:2022nwu}
(see eqn. 4.21), the purity at the end of inflation for a CMB mode
is (using the observed power spectrum),
\[
\gamma_{k_{\rm CMB}}^{{\rm end}}\simeq2.1\times10^{-35}\left(\frac{M_{{\rm P}}}{\rho_{{\rm inf}}^{1/4}}\right)^{4}
\]
which gives an entropy per CMB mode of $O(10)$ from the above formula. In any case the period of inflation is finite and in general the FRW phase takes over long before $\eta$ goes to zero\footnote{Note on duration of inflation: Assume that $H$ is constant during
inflation (approximately true). We have, integrating the definition
$H$ and taking (with no loss of generality) the end of inflation
to be $t=0$ (where $t$ is proper time), $a(t)=a_{e}e^{Ht},\,t\le0$,
$a_{e}$ being the scale factor at the end of inflation. As we mentioned
in the introduction we are working in a cut off theory, and in any
case the inflationary phase is not expected to extend all the way
to the cosmological singularity (i.e. $a=0)$. So let us take the
onset of inflation to be at scale factor $a_{i}$. Integrating the
definition of conformal time $d\eta/dt=1/a(t)$ we have, $\eta-\eta_{i}=-1/\left(a(t)H\right)+1/\left(a_{i}H\right)$.
Taking $\eta_{i}=-1/a_{i}H$ we have the usual expression $\eta=-1/\left(a(t)H\right)$.
The duration of inflation is in any case finite, $\eta_{e}-\eta_{i}=-1/\left(a_{e}H\right)+1/\left(a_{i}H\right)$.
So since during inflation $a$ is neither zero nor infinite our calculations
have neither a UV nor an IR divergence.\label{etatrelation}}. 


Let us now consider how far these  calculations (albeit in a
Gaussian state) can be replaced by simply considering the diagonal
element 
\begin{equation}
P\left(\phi\right):=\prod_{k}\langle\phi_{k}|{\rho}_{{\rm red}}|\phi{}_{k}\rangle=\prod_{k}P\left(\phi_{k}\right),\qquad P\left(\phi_{k}\right):=\sqrt{\frac{1}{2\pi\Sigma_{k11}}}\exp\left(-\frac{\phi_{k}^{2}}{2\Sigma_{11}}\right)\label{eq:Pphi}
\end{equation}
 which is what would give rise to the stochastic formulation treated
in the previous section. The purity calculated from this is $\gamma_{k}^{{\rm diag}}={\rm tr}\left({\rho}^{\rm diag}\right)^{2}=\frac{1}{2\sqrt{\pi\Sigma_{11}}}$.
Similarly the entropy calculated from this (essentially what one might
naively expect from decoherence theory as discussed in the previous
section), is $S^{{\rm diag}}=\ln\left(2\pi\Sigma_{11}\right)+\frac{1}{2}$.
These differ significantly from the exact calculation, in particular
there is no dependence on $\Sigma_{22}$ in this ``approximate''
calculation. In fact as we will see this difference is such that at
late times there is simply no sense in which this calculation can
approximate the exact calculation.

To see this and understand what goes wrong with the approximation
of dropping the off-diagonal terms in the calculation of the previous
section, it is convenient to change variables to $\phi_{{\rm s}}=\frac{1}{2}\left(\phi+\phi'\right),\,\phi_{{\rm a}}=\phi-\phi'$
in which case one has (dropping the second factor in \eqref{eq:Gaussian}
and the sub and superscripts $k,R,I$ on $\phi_{{\rm s,a}}$)
\begin{equation}
\langle\phi|{\rho}_{{\rm red}}|\phi'\rangle=\frac{1}{\sqrt{2\pi\Sigma_{11}}}\exp\left[-\frac{1}{2\Sigma_{11}}\phi_{s}^{2}-\frac{\det\Sigma}{2\Sigma_{11}}\phi_{{\rm a}}^{2}+i\frac{\Sigma_{12}}{\Sigma_{11}}\phi_{{\rm s}}\phi_{{\rm a}}\right]\label{eq:saGaussian}
\end{equation}
This shows that at late times, when it is expected that $\det\Sigma\rightarrow\infty$
the dependence on $\phi_{{\rm a}}=\phi-\phi'$ i.e. the off-diagonals
in $\rho$ are exponentially suppressed. However it should be noted
that (contrary to the claim after eqn. 4.7 of \citep{Burgess:2022nwu})
it is not a delta function (of $\phi-\phi'$) which would give just
the diagonal piece. In fact it is clear from the second term in the
exponent that what should go to infinity in the coefficient of $\phi_{{\rm a}}^{2}$ is
$\frac{\det\Sigma}{\Sigma_{11}}\sim\Sigma_{22}$. This kind of behavior
is not seen in for example spin systems (where most of the arguments
about decoherence are made). It is the difference between the continuum
system studied here from discrete systems that accounts for the fact
that the reasoning that led to the last line of \eqref{eq:Stodiagonal}
is incorrect.

In fact in the limit $\Sigma_{22}\rightarrow\infty$ as pointed out
below \eqref{eq:GaussianEntropy} the purity goes to zero and the
entropy goes to infinity. However when computed using \eqref{eq:Pphi}
the purity remains non-zero and the entropy remains finite since they
are both independent of $\Sigma_{22}$. This means that there is no
way that $P\left(\phi\right)$ can approximate (a Gaussian) density
matrix at late times. 

\subsection{The second law in cosmology for a Gaussian state}

Nevertheless, it is possible to derive a second law for accelerated
cosmologies for these Gaussian states. We again use the set up of
\citep{Burgess:2022nwu} which (apart from using the above Gaussian
representation of $\rho$ and the assumptions that lead to the Lindblad
evolution of $\rho_{{\rm red}}$), argues that the dominant
interaction of the curvature fluctuation $\zeta$ \footnote{Here we are identifying
$\zeta$ with our $\phi$} is of the form $\zeta\partial\zeta\partial\zeta$
and is therefore linear in the super-horizon mode (while being quadratic
in the sub-horizon modes). In this case it is shown there that (see
eqn. (4.8) of that reference)
\begin{equation}
\frac{\partial}{\partial\eta}\det\Sigma=2{\mathrm{Re}}\,   \left[{\cal F}_{k}(\eta,\eta_{0})\right]\Sigma_{11},\label{eq:detSigmadot}
\end{equation}
 and (see eqn. (3.37) and the sentence below it in \citep{Burgess:2022nwu} ), as pointed out in
this reference ${\mathrm{Re}}\,   \left[{\cal F}_{k}(\eta,\eta_{0})\right]$is expected
to be positive on general grounds\footnote{This of course would have followed from the properties of the Lindblad
equation, in particular the necessity of preserving the trace and
 the positivity of the reduced density matrix. However here we
do not have strict Markovianity so this is not guaranteed. For a detailed
discussion of this including a generalization of the arguments of
    \citep{Burgess:2022nwu}, see \citep{Lopez:2025arw}.}. Explicitly in the model that is investigated in detail there,
\[
{\mathrm{Re}}\,   \left[{\cal F}_{k}(\eta,\eta_{0})\right]\simeq\frac{\epsilon H^{2}k^{2}}{1024\pi^{2}M_{P}^{2}}\left\{ \frac{20\pi}{\left(-k\eta\right)^{2}}+\ldots\right\} ,\qquad{\rm for}\qquad -k\eta\ll1.
\]
This expression (calculated in perturbation theory) will break down
at late times but in the following we will only make use of the positivity
of ${\mathrm{Re}}\,   \left[{\cal F}_{k}(\eta,\eta_{0})\right]$ which is expected
to hold non-perturbatively.

Hence
\begin{equation}
\frac{\partial}{\partial\eta}\det\Sigma>0,\qquad {\rm for}\qquad -k\eta\ll1.\label{eq:detSigmadot2}
\end{equation}
So it seems that in this set up $\det\Sigma$ increases without bound
for late times and hence the purity appears to go to zero. However
as pointed out in the paragraph after eqn. \eqref{eq:GaussianEntropy} in the previous subsection,
purity at the end of inflation for a CMB mode is small but non-zero\footnote{Of course as pointed out in footnote \ref{etatrelation} since the period of inflation is finite the purity never reaches zero before the end of inflation.}.

Now consider the time variation of \eqref{eq:GaussianEntropy} at
fixed cut-off. 
\begin{equation}
\frac{dS^{(k)}}{d\eta}\Bigg |_{\Lambda}=\frac{dS^{(k)}}{d\gamma}\frac{d\gamma}{d\eta}\Bigg |_{\Lambda}=\left(-\frac{1}{2\gamma_{k}^{2}}\ln\left(\frac{1+\gamma_{k}}{1-\gamma_{k}}\right)\right)\left(-\frac{1}{4\left(\det\Sigma\right)^{3/2}}\frac{\partial}{\partial\eta}\det\Sigma\Bigg |_{\Lambda}\right)>0,\label{eq:SkdotfixedL}
\end{equation}
where the last relation follows from \eqref{eq:detSigmadot2}. The
total entropy (just considering the $R$ factor in \eqref{eq:Gaussian})
is given by (switching now to continuum k-space),
\[
S(\eta,\Lambda)=V_{\rm space}\frac{1}{\left(2\pi\right)^{2}}\int_{0}^{\Lambda}k^2dkS^{(k)}.
\]
 The total time derivative of the entropy is,
\begin{equation}
\frac{dS}{d\eta}=V_{\rm space}\frac{1}{\left(2\pi\right)^{2}}\left( \int^{\Lambda}k^2dk\frac{dS^{(k)}}{d\eta}\Bigg |_{\Lambda}+S^{(\Lambda )}\Lambda^2\frac{d\Lambda}{d\eta}\right) .\label{eq:totaltimederivative}
\end{equation}
The first term is positive from eqn. \eqref{eq:SkdotfixedL} and since
with $\Lambda=aH$, $$\frac{d\Lambda}{d\eta}=a{\ddot{a}}\qquad {\rm then}\qquad  \ddot a>0\implies \dot S>0,$$ i.e. the
second term is also positive as long as the universe is accelerating.
Thus we have shown that during accelerating periods (such as during
inflation both primordial and current) the quantum entropy is increasing.

In order to avoid confusion let us stress that the Lindblad equation argument is only used in the first term of \eqref{eq:totaltimederivative}. In other words the Linblad equation computes $\partial\rho/\partial\eta|_{\Lambda}$ i.e. with a fixed cutoff. The time dependence of the cutoff is taken into account in the second term of \eqref{eq:totaltimederivative}.

Note added in version 2. After this paper was uploaded to ArXiv, an article~\cite{Burgess:2025dwm} by the first three authors of reference \cite{{Burgess:2022nwu}} appeared  in which it was argued that the dominant interaction is different from the one considered in  \cite{{Burgess:2022nwu}}. However while that will affect explicit expressions such as the equation below \eqref{eq:detSigmadot}, it must still be linear in the system field as argued in these references. So our discussion which depends only on the general properties of these calculations will not be affected. Indeed the discussion in the next section just depends on there being just one dominant interaction during inflation, a situation that obataines in both these references.

\subsection{Second law in cosmology beyond the Gaussian approximation}

Here we give an argument for \eqref{eq:SkdotfixedL} which does not
rely of using a Gaussian state for the reduced density matrix. Instead
we exploit the fact that, as pointed out in \citep{Burgess:2022nwu},
effectively one needs to consider only one term in the interaction
Hamiltonian between the system and environment modes. One has (see
eqn. 3.1 of \citep{Burgess:2022nwu}) with $v$ a superhorizon (system)
field and $B$ a subhorizon (environment) (composite) field\footnote{The following arguments assume that this interaction is dominant. A similar argument can also be made when one other interaction dominates as discussed in \citep{Lopez:2025arw}.},
\begin{equation}
H_{{\rm int}}\left(\eta\right)=G\left(\eta\right)\int d^{3}xv\left(\eta,{\bf x}\right)\otimes B\left(\eta,{\bf x}\right).\label{eq:Hint}
\end{equation}
To show purity loss the authors then proceed to derive a Linblad equation.
The derivation given in \citep{Burgess:2022nwu} in particular the
positivity of the relevant coefficient ${\rm Re}\left[{\cal F}_{k}(\eta,\eta_{0})\right]$
and its independence from $\eta_{0}$ depends on the details of a
long laborious calculation given in that reference. Instead let us
assume (as is often done in deriving the Lindblad equation) that the time
scale over which the environment correlation function decays is much
shorter than the time scale over which the density matrix changes (for
a fixed cut-off). Let us look at this procedure in more detail. We work
to second order  in the above interaction in the interaction
picture and ignore the commutator term $[H_{{\rm int}},\rho]$
which does not contribute to $dS/d\eta$.  We get (note below we have
put $\rho_{red}\rightarrow\rho$ so following \citep{Burgess:2022nwu}
the total initial state is $\rho\otimes|0_{B}\rangle\langle0_{B}|$ the second factor
being the Bunch-Davies vacuum for the high frequency modes), 
\begin{align}
\frac{\partial\rho}{\partial\eta} & \simeq-\int d^{3}x\int d^{3}x'\int_{\eta_{0}}^{\eta}d\eta'G(\eta)G(\eta')\nonumber \\
 & \left( \left[v\left(\eta,{\bf x}\right),v\left(\eta',{\bf x}'\right)\rho\left(\eta'\right)\right]C_{B}\left(\eta,\eta';{\bf x-x'}\right)+\left[\rho\left(\eta'\right)v\left(\eta',{\bf x'}\right),v\left(\eta,{\bf x}\right)\right]C^*_{B}\left(\eta,\eta';{\bf x-x'}\right)\right) .\label{eq:drhodeta}
\end{align}
Here (with $\Delta B:=B-\langle 0_{B}|B|0_{B}\rangle$), $C_{B}\left(\eta,\eta';{\bf x-x'}\right):=\langle 0_{B}|\Delta B\left(\eta,{\bf x}\right)\Delta B\left(\eta',{\bf x}'\right)|0_{B}\rangle$.
Note that the Hermiticity of $B$ implies that 
\begin{equation}
C_{B}^{*}\left(\eta,\eta';{\bf x-x'}\right)=C_{B}\left(\eta',\eta;{\bf x'-x}\right)\label{eq:CstarC}
\end{equation}

Now we make the crucial assumption that is often made in deriving
the Lindblad equation from the unitary evolution of the total state,
namely that the time correlation function of the environment modes
decays on a time scale much shorter that the time scale over which
$\rho$ changes. Effectively we are going to assume that the $\eta'$
integral collapses to a contribution at the upper limit. In other
words we approximate the integral by assuming that $C$ is effectively
proportional to $\delta(\eta-\eta')$. Writing $C_{B}\left(\eta,\eta';{\bf x-x'}\right)\simeq\delta(\eta-\eta')C\left(x-x'\right),$ and since the $[B\left(|\eta,{\bf x}\right),B\left(|\eta',{\bf x}'\right)]=0$
for space like separations (which is the case in our approximation
where $\eta'-\eta\simeq0$), we have from \eqref{eq:CstarC} 
\begin{equation}
C_{B}^{*}\left({\bf x-x'}\right)=C_{B}\left({\bf x'-x}\right)=C_{B}\left({\bf x-x'}\right),\label{eq:CrealSymm}
\end{equation}
i.e. $C$ is a real symmetric function of ${\bf x'-x}$. After making
these approximations and some manipulation  \eqref{eq:drhodeta}
becomes\footnote{Since effectively the dominant contribution to the integral comes from the end point of the integration only half the peak is relevant so we have approximated this by writing 
$\int_{\eta_0}^{\eta}\delta(\eta'-\eta)f(\eta')=1/2f(\eta)$. }
\begin{equation}
\frac{\partial\rho}{\partial\eta}  \simeq-\frac{1}{2}\int d^{3}x\int d^{3}x'G^{2}(\eta)\left( \left\{ v\left(\eta,{\bf x}\right)v\left(\eta,{\bf x}'\right),\rho(\eta)\right\} -2v\left(\eta,{\bf x}\right)\rho(\eta)v\left(\eta,{\bf x}'\right)\right) C_{B}\left({\bf x-x'}\right)\label{eq:RHSanticommform}
\end{equation}
This has the form of a time-dependent Lindblad equation~\cite{Martin:2018zbe}. From here we can compute the variation of the quantum entropy $S=-{ \rm tr}\, \rho\, {\rm ln}\rho$ given by  $\frac{dS}{d\eta}=-{\rm tr}\, \frac{d\rho}{d\eta}\ln\rho$. 

Now let us assume that the background is spatially compact so that
the density matrix can be expanded in a  eigenbasis labelled
by $\alpha,\beta$ etc. For instance in the absence of interaction
(assumed to be the case in the far past) we have $\rho=\prod_{k_{\alpha}>\Lambda}\int_{\phi_{k}}\langle\phi_{k_{\alpha}}|\left\{ \phi_{k'}\right\} \rangle\langle\{\phi_{k'}\}|\phi_{k_{\alpha}}\rangle$,
where the $k_{\alpha}$'s take discrete set of values taking space to be compact as before. With time evolution
this basis will of course get transformed, but given that the density
matrix for the system is still Hermitian there will still be some
basis in which $\rho$ will be diagonalized. To simplify the notation
we label this basis $\left\{ |\alpha\rangle\right\} $ and take it to be discrete with $<\alpha |\beta >=\delta_{\alpha\beta}$. So $\rho=\sum_{\alpha}p_{\alpha}|\alpha\rangle\langle\alpha|,$
$\sum_{\alpha}p_{\alpha}=1,\,0<p_{\alpha}<1$.

Using \eqref{eq:RHSanticommform} and evaluating the trace in this
eigenbasis we get
\begin{align*}
\frac{\partial S}{\partial\eta}\vert_{\Lambda} & \simeq\int d^{3}x\int d^{3}x'G^{2}\left(\eta\right)\left(\sum_{\alpha\beta}(v(\eta,{\bf x})v(\eta,{\bf x}'))_{\alpha\beta}\left(p_{\alpha}\ln\rho_{\alpha}\right)\delta_{\alpha\beta}+p_{\alpha}\left((v(\eta,{\bf x})v(\eta,{\bf x}')\right)_{\alpha\beta}\left(\ln p_{\alpha}\right)\delta_{\beta\alpha}\right.\\
 & - \left. 2\left(v(\eta,{\bf x})\right)_{\alpha\beta}p_{\beta}\left(v(\eta,{\bf x}')\right)_{\beta\alpha}\ln p_{\alpha}C_{B}\left({\bf x-x'}\right)\vphantom{\sum_{\alpha\beta}}\right) \\
 &= \int d^{3}x\int d^{3}x'G^{2}\left(\eta\right)\sum_{\alpha,\beta}\left\{ v_{\alpha\beta}\left(\eta,{\bf x}\right)v_{\alpha\beta}^{*}\left(\eta,{\bf x}'\right)C_{B}\left({\bf x-x'}\right)\left(p_{\alpha}-p_{\beta}\right)\ln p_{\alpha}\right\} .
\end{align*}
In the last step we used the hermiticity of $v$, i.e. $v_{\beta\alpha}=v_{\alpha\beta}^{*}$. 
Now note that 
\begin{align}
\sigma_{\alpha\beta} & :=\int d^{3}xd^{3}x'v_{\alpha\beta}\left(\eta,{\bf x}\right)v_{\alpha\beta}^{*}\left(\eta,{\bf x}'\right)C_{B}\left({\bf x-x'}\right)=\int d^{3}xd^{3}x'v_{\beta\alpha}^{*}\left(\eta,{\bf x}\right)v_{\beta\alpha}\left(\eta,{\bf x}'\right)C_{B}\left({\bf x-x'}\right)\nonumber \\
 & =\int d^{3}xd^{3}x'v_{\beta\alpha}^{*}\left(\eta,{\bf x'}\right)v_{\beta\alpha}\left(\eta,{\bf x}\right)C_{B}\left({\bf x'-x}\right)=\int d^{3}xd^{3}x'v_{\beta\alpha}\left(\eta,{\bf x}\right)v_{\beta\alpha}^{*}\left(\eta,{\bf x'}\right)C_{B}\left({\bf x-x}'\right)\nonumber \\
 & =\sigma_{\beta\alpha}\label{eq:sigma}
\end{align}
In the first line above we've used the Hermiticity of $v$ and in
the second line the symmetry \eqref{eq:CrealSymm}. Also we have
\begin{equation}
\sigma_{\alpha\beta}={\Bigg\langle}\int_{x}v_{\alpha\beta}({\bf x})\Delta B({\bf x})\int_{x'}v_{\alpha\beta}^{*}({\bf x}')\Delta B({\bf x}')\,{\Bigg\rangle}_{B}=\Bigg\langle\,\Bigg\vert\int_{x}v_{\alpha\beta}({\bf x})\Delta B({\bf x})\, \Bigg\vert^{2}\, \,{\Bigg\rangle}_{B}\ge0.\label{eq:sigmapositvie}
\end{equation}
Hence we have \footnote{The rest of this argument follows one given by Weinberg \citep{Weinberg:qm} for
the QM case under the assumption that the Lindblad operator is Hermitian.
Here given that only one interaction is dominant, as argued for in
\citep{Burgess:2022nwu}, we are effectively in a similar situation. Note that in contrast to the discussion in  \citep{Weinberg:qm} we are actually in a field theoretic situation and it
was crucial to observe that once time locality was assumed the environment operators commute since they are at space-like separation. } 
\begin{align}
\frac{\partial S}{\partial\eta}\vert_{\Lambda} & \simeq G^{2}\sum_{\alpha\beta}\sigma_{\alpha\beta}\left(p_{\alpha}-p_{\beta}\right)\ln p_{\alpha}=\sum_{\alpha\beta}\sigma_{\beta\alpha}\left(p_{\beta}-p_{\alpha}\right)\ln p_{\beta}\nonumber \\
 & =G^{2}\sum_{\alpha\beta}\sigma_{\alpha\beta}\left(p_{\beta}-p_{\alpha}\right)\ln p_{\beta}\nonumber \\
 & =G^{2}\sum_{\alpha\beta}\sigma_{\alpha\beta}\left(p_{\alpha}-p_{\beta}\right)\left(\ln p_{\alpha}-\ln p_{\beta}\right)\ge0.\label{eq:2ndLaw}
\end{align}
In the second line above we used \eqref{eq:sigma} and in the last
line the positive semi-definiteness of $\sigma$, \eqref{eq:sigmapositvie}. Also the quantity $\left(p_{\alpha}-p_{\beta}\right)\left(\ln p_{\alpha}-\ln p_{\beta}\right)$ is positive since the log is a monotonic function of its argument. We then conclude that entropy increases monotonically.

The above argument was made however for a fixed separation between the system and the environment i.e.  for $\Lambda$ fixed. In other words, taking  the entropy $S(\eta,\Lambda)=V_{\rm space}\frac{1}{\left(2\pi\right)^{2}}\int_{0}^{\Lambda}k^2dkS^{(k)} $ we computed the time derivative keeping the cut-off fixed. Thus we have an alternative argument to that given in \citep{Burgess:2022nwu} for the first term of equation \eqref{eq:dSbydt}. The positivity of the second term follows as before when $\ddot a>0$.


\section{Discussion}


\subsection{Von Neumann entropy and thermal entropy in cosmology}

The entropy discussed above, with non-negative time derivative during accelerated expansion arising from decoherence,  is the von Neumann entropy. In contrast, the standard entropy budget of the universe refers to the thermodynamic entropy associated with black holes, photons, neutrinos, etc., each weakly coupled system typically in thermal equilibrium and with entropies  largely constant over time. This thermodynamic entropy is usually assumed to be generated at the end of inflation, when the inflaton's potential energy $V$ was converted into kinetic energy and then into standard model particles at a temperature roughly $T_{{\rm bigbang}}\sim V^{1/4}$.  The post-inflationary universe then evolves with  constant entropy per comoving volume $S_{{\rm thermo}}\sim a^{3}T^{3}$.

In the original ``entropy problem'' the question was how to account
for the large entropy in the observable universe, dominated by the CMB and estimated to be $S\sim10^{89}$. However, black hole formation after inflation introduces  a larger entropy, by using the Beckenstein-Hawking formula, $S_{{\rm bh}}\sim10^{102}$ and the entropy coming from the cosmological horizon could be even larger, potentially contributing $S\sim10^{122}$. Even ignoring the latter, we face two key questions:

\begin{enumerate}
\item How to account for the large entropy in black holes from post inflationary
physics?
\item What is the relation between these thermal entropies and the von Neumann
entropy coming from decoherence of inflationary fluctuations?
\end{enumerate}

Addressing question 1, black hole entropy reflects the degeneracy of energy eigenstates and is the maximum of both Gibbs and von Neumann entropies. Yet since gravitational collapse is unitary, this entropy can only be an upper bound on that of the initial state. Ejected matter from the collapsing star may be entangled with infalling matter, but this may not alter the total entropy of the universe, which remains conserved assuming that total state evolves unitarily.

For question 2, we note that the von Neumann entropy of the CMB modes  increases 
during accelerated expansion (inflation and the current dark energy phase), due to 
entanglement between sub and super horizon modes. During the intermediate decelerating radiation/matter dominated phases, the total time derivative of the entropy can still be non-negative due to contributions from terms independent of  $\ddot{a}$, i.e. the first term of equation \eqref{eq:dSbydt} may dominate the second. However the usual adiabatic assumption in this regime implies that entropy per co-moving volume is constant. Of course this refers to thermodynamic entropy, but it is not clear to us how to  identify it with the von Neumann entropy of the fluctuations\footnote{See for instance the review \cite{CALABRESE201831} for a pedagogical discussion of this connection.}.

Since the total state of the inflaton
fluctuations is assumed to be in a pure state, the entropies of system and environment remain equal. After reentry, the CMB modes within
the horizon constitute the ``system''.
Once all the modes that result
through gravitational collapse to form the present structures have
reentered the horizon during the FLRW phase, the von Neumann entropy
of these (observable) modes should remain approximately constant since the entanglement with the deep UV modes can probably be ignored.

We must also
account for the homogeneous background that is seen to fit the radiation
spectrum of a perfect black body at a temperature of about $2.7{\rm K}$ with entropy  $10^{89}$.
 The standard picture posits thermalisation at the end of inflation, converting the inflaton's  energy into a thermal bath of particles (see for instance \cite{Kofman:1997yn})\footnote{Note that increase of entropy during inflation has also been considered assuming the geometric expression for the entropy in terms of the area of the cosmological horizon $S\propto 1/H^2$. See for instance \cite{Bousso:2006ge,Arkani-Hamed:2007ryv}.}. However, from a fundamental viewpoint, the universe is a closed quantum system evolving unitarily, which cannot strictly lead to thermalisation.
 
 Nonetheless, as shown in quantum many-body and field theory studies (e.g.,  \citep{DAlessio:2015qtq, Berges:2004yj}), unitary evolution can mimic thermalisation over long timescales, producing a behaviour indistinguishable from thermal equilibrium, despite retaining all information. As such the actual von Neumann entropy of the background remains constant, while the system appears thermal, at a temperature that, according to the above
references, is to be identified with the average momentum of the quanta
of the state at late times.

Thus from a fundamental point of view there is no entropy increase
at the end of inflation when the potential energy of the inflaton is converted into the energy of (say) standard model particles.. The apparent thermal state that is observed is simply
a consequence of the late time behaviour of the quantum evolution
being indistinguishable from a thermal state. If these arguments are
correct then the only reason that $dS/dt$ may be positive (as opposed
to zero) are the arguments given in the previous section\footnote{It is then conceivable to consider the early universe to start in a pure state, and then the only apparent increase in entropy is the limitation of having no access to states outside the horizon. Thinking in terms of a global observer, both the total entropy and the total energy of our universe may just be zero, consistent with creation from nothing scenarios \cite{Vilenkin:1982de,Hartle:1983ai}.}.  Clearly,  these matters need further understanding.

\subsection{General Conclusions}

Summarizing, in this note we have studied decoherence and entropy increase in a cosmological set-up. 
While we do not claim to resolve all the fundamental questions regarding entropy in our universe, we have contented ourselves to address  a  concrete question by considering the entropy due to the inflationary CMB modes, for which there is a natural separation between system and environment, given by sub and super horizon modes respectively. Since this separation is time dependent, the corresponding entanglement entropy and its time evolution depends on the time evolution of the corresponding cut off.

In order to gain some understanding we compared the cosmological system with simple discrete bi-partite systems for which the entanglement entropy can be computed and entropy increase is related to quantum decoherence. We find differences between the systems. In particular the role of the off-diagonal terms of the reduced density matrix is different in both cases, decaying in the discrete system but could not be neglected in the  cosmological set up. However, besides this difference we still concluded that the entropy can increase monotonically in both cases. In cosmology, this is guaranteed only in periods in which the universe is accelerating such as in early universe inflation and current dark energy domination. In non-accelerating periods the time evolution of the entropy is given by the sum of two terms in \eqref{eq:dSbydt} that can be of any sign. 

Our approach may be compared with other discussions of entropy and density matrix in cosmological set-ups. In particular, the calculation of the entanglement entropy of field theory in de Sitter space in \cite{Maldacena:2012xp} and the related recent work in \cite{Ivo:2024ill} which consider  the density matrix for a system limited by the cosmological horizon. However they consider the density matrix for a subregion in a slice of physical space rather than the momentum space considered here.  In our approach, the UV divergence of the system is irrelevant since we work in an effective field theory with a UV cut-off $\Lambda_{UV}\sim M_P$ with $\Lambda(t)=aH\ll M_P$. In some sense separating system and environment in terms of long and short momentum modes, is close to a renormalisation group flow analysis and the entropy increase may be speculated to correspond to a sort of $c$-theorem~\cite{Balasubramanian:2011wt}. 
Whereas the tracing out in coordinate spacetime corresponds to tracing over the degrees of freedom in a physical region that is not causally connected with the observer. In coordinate spacetime there is also the standard short-distance divergent contribution to the entropy but it  has to be resolved by imposing a cut-off. It may be interesting to better understand the relation between these two approaches.

It is important to notice that we concentrated only on the contribution to the entropy of the CMB modes. We argue that other contributions may correspond to unitary evolutions for which Liouville's theorem is at work. We leave a more comprehensive analysis including different types of entropies for a future publication.

\section*{Acknowledgements}
We acknowledge interesting conversations with Ahmed Almheiri, Cliff Burgess, Thomas Colas, Paolo Creminelli, Steve Gratton, Greg Kaplanek, Simon Lin, Juan Maldacena, Daniel Mata-Pacheco, Mehrdad Mirbabayi, Shoy Ouseph, Enrico Pajer, Pellegrino Piantadosi and Luis Zapata. We thank specially to Thomas Colas, Greg Kaplanek and  Enrico Pajer for a careful reading of the manuscript.  SC thanks the hospitality  of New York University Abu Dhabi where part of this work was carried. SC is supported in
part by the STFC Consolidated Grants ST/T000791/1 and ST/X000575/1 and by a Simons
Investigator award 690508. The research of FQ is funded by a NYUAD research grant.


\begin{appendix}
\section{Details on the path integral derivations}

\label{Appendix:CoarseGraining}
In this appendix we provide a field theoretical approach towards computing the purity and entropy in an open system.

Let us start our discussion by considering a scalar field $\phi$ with an action given by,
\begin{align}
    S[\phi]=\int d^4x \sqrt{-g}\left(-(\partial_\mu\phi)^2+\mathcal{L}_{\mathrm{int}}[\phi]\right)
\end{align}
Now we will split the action between long and short modes as
\begin{align}
    \phi=\phi^{l}+\phi^s
\end{align}
If we expand the action then we have that,
\begin{align}
    S[\phi_l+\phi_s]=S^l[\phi_l]+S^s[\phi_s]+\Delta S[\phi_l,\phi_s]
\end{align}
where 
\begin{align}
    S^{l}[\phi_l]=-\int_{k\leq \Lambda} \int dt \sqrt{-g}\left((\partial_\mu\phi_l)^2+\mathcal{L}^l[\phi_l]\right)
\end{align}
and similarly for $S^s$. In what follows we will assume that we can expand the action in powers of $\phi_s$ as follows
\begin{align}
    S=\int d^3x\int dt \sqrt{-g} \left(-\frac{1}{2}(\partial_\mu\phi_l)^2-\frac{1}{2}(\partial_\mu\phi_s)^2+\mathcal{L}[\phi_l]+\frac{\partial \mathcal{L}[\phi_l]}{\partial\phi_l}\phi^s-\frac{1}{2}\frac{\partial^2 \mathcal{L}[\phi_l]}{\partial\phi_l^{2}}
\phi_s^2\right)
\label{slexpansion}\end{align}
\paragraph{Schwinger-Keldysh contour}

We want to define  a density matrix on the Schwinger-Keldysh contour given by first evolving the field from the Bunch-Davies vacuum in the past, up to some finite time $t_0$ in the future and then evolving it back to the vacuum. This implies that the density functional is given by,
\begin{align}
    \rho[\phi_+,\phi_-]=\int_{\mathrm{BD}}^{\Phi_+(t_0)=\phi_+} D\Phi_+\int_{\mathrm{BD}}^{\Phi_-(t_0)=\phi_-} D\Phi_- e^{iS[\Phi_+]-iS[\Phi_-]}
\end{align}
The path integral can be solved by using the saddle point approximation around the classical solution $\phi^{\mathrm{cl}}$, we then find,
\begin{align}
     \rho[\phi_+,\phi_-]=e^{iS[\phi^{\mathrm{cl}}_+]-iS[\phi^{\mathrm{cl}}_-]}
\end{align}
where the classical solution is usually written in terms of the \textit{bulk-to-boundary} $K(k,t)$ and \textit{bulk-to-bulk} $G(k,t,t')$ as~\cite{Cespedes:2023aal},
\begin{align}
  \Phi_{\mathrm{cl}}(k,t)= K(k,t)\phi_{\mathrm{cl}}+i \int dt' G(k;t,t')\frac{\delta S_{\mathrm{int}}}{\delta\Phi_{\mathrm{cl}}}
  \end{align}
both propagators need to satisfy the boundary conditions dictated by the path integral such that $\Phi(k,t_0)=\phi_{\mathrm{cl}}$ and that at $\lim_{t\to-\infty}$ the field is at the vacuum. Notice that $G(k,t,t')$ is the Green function satisfying the boundary conditions imposed by the path integral.
\begin{align}
    (\Box-m^2)G(k;t,t')=\frac{i}{\sqrt{-g}}\delta(t-t'),\qquad \lim_{t\to-\infty,t_0}G(k;t,t')=0
\end{align}In terms of the mode function $f_{k}(t)$ the propagators are given by
\begin{align}
    K(k,t)&= \frac{f_{k}(t)}{f_{k}(t_0)},\nonumber\\
    G(k;t,t')&=f^*_k(t)f_k(t')\theta(t-t')+f^*_k(t')f_k(t)\theta(t'-t)-\frac{f^*_k(t_0)}{f_k(t_0)}f_k(t)f_k(t')
\end{align}

\paragraph{Reduced density matrix}
We are interested in integrating out over modes shorter than some scale  $\Lambda$ and compute a reduced density matrix. This is defined as,
\begin{align}
    \rho_{\mathrm{red}}[\phi^l_+,\phi^l_-]=\mathrm{Tr}_{k\geq\Lambda}\rho[\phi_+,\phi_-]
\end{align}
Operationally and in perturbation theory this can be done by expanding the action in powers of small and large fields and then integrate over the short wavelength fields. It follows that,
\begin{align}
    \rho_{\mathrm{red}}[\phi^l_+,\phi^l_-]&=\int_{\mathrm{BD}}^{\Phi^l_+(t_0)=\phi^l_+} D\Phi^l_+\int_{\mathrm{BD}}^{\Phi^l_-(t_0)=\phi^l_-} D\Phi^l_+\int d\phi^s\int_{\mathrm{BD}}^{\Phi^s_+(t_0)=\phi^s} D\Phi^s_+ \int_{\mathrm{BD}}^{\Phi^s_-(t_0)=\phi^s} D\Phi_-^s\nonumber\\
    &\qquad \times e^{iS[\Phi_+^l+\Phi_+^s]-iS[\Phi_-^l+\Phi_-^s]}
\end{align}
Let us now assume that we can write down the classical solution for the long/short wavelength modes as,
\begin{align}
    \Phi^l_{\mathrm{cl}}(k,t)= K^l(k,t)\phi^l_{\mathrm{cl}}+i \int dt' G^l(k;t,t')\frac{\delta S_{\mathrm{int}}}{\delta\Phi^l_{\mathrm{cl}}}
\end{align}
and where the propagators are defined as,
\begin{align}
    K^l(k,t)&=\int d^3 xe^{i k x}\int_{q<\Lambda}e^{iqx}K(k,t)\nonumber\\
    G^l(k;t,t;)&=\int d^3 xe^{i k x}\int_{q<\Lambda}e^{iqx}G(k;t,t')
\end{align}
and analogous for the short wavelength modes. Let us first consider the part of the action that depends only on the short modes. If we use the expansion from \eqref{slexpansion} and assume that   $\Delta\mathcal{L}(\phi^l,\phi^s)=g\phi^l\phi^s$, we get:
\begin{small}
\begin{align}
    S[\Phi_+]- S[\Phi_-]&\supset \int_{\bm{k}}\int d\eta\sqrt{-g}\left(-(\partial_\mu\Phi_+^s)^2-\Delta\mathcal{L}(\Phi_+^l,\Phi_+^s)\right)-\int_{\bm{k}}\int d\eta\sqrt{-g}\left(-(\partial_\mu\Phi_-^s)^2-\Delta\mathcal{L}(\Phi_-^l,\Phi_-^s)\right)\nonumber\\
    &=\int_{\bm{k}}i\psi_s(\phi_+^{s})^2-g\int_{\bm{k}}\int d\eta\sqrt{-g} K^s(k,\eta)\phi^s_+\Phi_+^l(\eta)-\int_{\bm{k}}i\psi_s^*(\phi_-^s)^2\nonumber \\
    &-g\int_{\bm{k}}\int d\eta\sqrt{-g} (K^s)^*(k,\eta)\phi^s_-\Phi_-^l(\eta)
    + \frac{ig^2}{2}\int _{\bm{k}}\int d\eta\sqrt{-g}\int d\eta'\sqrt{-g}\Phi_+^l(\eta)G^s(k;\eta,\eta')\Phi_+^l(\eta')\nonumber\\
    &- \frac{ig^2}{2}\int _{\bm{k}}\int d\eta\sqrt{-g}\int d\eta'\sqrt{-g}\Phi_-^l(\eta)(G^s)^*(k;\eta,\eta')\Phi_-^l(\eta')
\end{align}
\end{small}
where we have defined $\psi_s=i\frac{ d }{d\eta}K(k,\eta,\eta_0)\vert_{\eta=\eta_0}$. The action is unitary in the sense that there are no mixed terms between the two branches of the 
Schwinger-Keldysh contour. However, it is non-local, this is evident in the last two terms, which can be interpreted as non local interactions for the long mode $\Phi_l$.

This becomes more transparent upon integrating out the short mode $\phi_s$. Since we are computing the trace over the environment, we begin by identifying $\phi_+^s = \phi_-^s$. The resulting action is quadratic in $\phi_s$, and the Gaussian integration over $\phi_s$ yields:
\begin{align}
    \rho_{\mathrm{red}}[\phi^l_+,\phi^l_=]&=\int^{\phi^l_+} D\Phi^l_+\int^{\phi^l_-} D\Phi^l_- \exp\left(iS^l[\Phi_+^l]-iS^l[\Phi_-^l]\right.\nonumber\\
    &\left.-\frac{g^2}{2}\sum_{\pm}\int_{\bm{k}}\int d\eta\sqrt{-g}\int d\eta'\sqrt{-g}\Phi_{\pm}^l(\eta)G^s_{\pm\pm}(k;\eta,\eta')\Phi_\pm ^l(\eta')\right).
\end{align}
where we have used the following identities for the propagators, 
\begin{align}
    G^{\sigma}(k;t,t')+\frac{K^\sigma(k,t)K^\sigma(k,t)}{2\mathrm{Re}\psi^\sigma_2}&=G_{++}^\sigma(k;t,t')\\
     (G^{\sigma})^*(k;t,t')+\frac{K_\sigma^*(k,t)K^*_\sigma(k,t)}{2\mathrm{Re}\psi^\sigma_2}&=G_{--}^\sigma(k;t,t')\\
     \frac{K_\sigma(k,t)K^*_\sigma(k,t)}{2\mathrm{Re}\psi^\sigma_2}&=G_{+-}^\sigma(k;t,t')
     \label{props_indent}
\end{align}
The resulting theory is now non-unitary, as it contains terms that mix the two branches of the Schwinger-Keldysh  contour, such terms cannot arise from evolution under a local Hamiltonian. The last term in particular represents the non-local influence functional, which encapsulates the effect of the environment on the system. Because it is also non-local in time, the dynamics are inherently non-Markovian.
However, when computing the reduced density matrix at a finite time, we can effectively bypass this complication. By evaluating the path integral on the classical saddle point configuration, the reduced density matrix takes the form:
\begin{align}
    \rho_{\mathrm{red}
    }[\phi_+^l,\phi_-^l]&=\exp\left[-\int_{\bm{k}}\psi_l(\phi_+^l)^2-\int_{\bm{k}}\psi_l^*(\phi_-^l)^2\right.\nonumber\\
    &\left.+\frac{g^2}{2}\sum_{\pm}\int_{\bm{k}}\left(\int d\eta\sqrt{-g}\int d\eta'\sqrt{-g}K_{\pm}^l(k,\eta)G^s_{\pm\pm}(k,\eta,\eta')K^l_{\pm}(k,\eta)\right)\phi^l_{\pm}\phi^l_{\pm}+\mathcal{O}(g^3)\right]
    \label{reducedrho_def}
\end{align}
If we define the following wavefunction coefficients,
\begin{align}
    \psi_W=\frac{g^2}{2}\int^{\eta_0} d\eta\sqrt{-g}\int^{\eta_0} d\eta'\sqrt{-g}K_{+}^l(k,\eta)G^s_{+-}(k;\eta,\eta')K^l_{-}(k,\eta),\\
    \psi_G=\frac{g^2}{2}\int^{\eta_0} d\eta\sqrt{-g}\int^{\eta_0} d\eta'\sqrt{-g}K_{+}^l(k,\eta)G^s_{++}(k;\eta,\eta')K^l_{+}(k,\eta).
   \label{def:wfncoeffs}
\end{align}
Then the reduced density matrix is simply given by,
\begin{align}
     \rho_{\mathrm{red}
    }[\phi_+^l,\phi_-^l]=\exp\left(-\int_{\bm{k}}(\psi_l+\psi_G)(\phi_+^l)^2-\int_{\bm{k}}(\psi_l^*+\psi_G^*)(\phi_-^l)^2+2\int \psi_W\phi_+^l\phi_-^l\right).
    \label{def:Wfn}
\end{align}
Notice that $\psi_W$ is positive since is the product of a function and its conjugate. This density matrix is equivalent to the Gaussian state considered in Eq.~\eqref{eq:Gaussian}. There the wavefunction coefficients are related to the covariance matrix. This is also the case in Eq.~\eqref{def:Wfn} as can be seen by computing the 2-point functions from the density matrix. Then the covariance matrix is given by,
\begin{align}
    \Sigma=\left(
    \begin{array}{cc}
       \frac{1}{\mathrm{Re}\psi_{l}+\mathrm{Re}\psi_{G}-\psi_W}  & -\frac{\mathrm{Im}\psi_{l}+\mathrm{Im}\psi_{G}}{2(\mathrm{Re}\psi_{l}+\mathrm{Re}\psi_{G}-\psi_W)} \\
    -\frac{\mathrm{Im}\psi_{l}+\mathrm{Im}\psi_{G}}{2(\mathrm{Re}\psi_{l}+\mathrm{Re}\psi_{G}-\psi_W)}    & \frac{\vert\psi_l+\psi_G\vert^2-\psi_W^2}{4(\mathrm{Re}\psi_{l}+\mathrm{Re}\psi_{G}-\psi_W)} 
    \end{array}\right),\qquad\mathrm{where} \ \psi^{(2)}=\psi_l+\psi_G
\end{align}
\paragraph{Purity} We can easily compute  the purity $\gamma=\mathrm{Tr}\rho^2$ from the expression above. All we need is to carefully take the trace after doubling the contour. This leads to,
\begin{align}
    \gamma&=\mathcal{N}\int d\phi_+\int d\phi_- e^{-\int_{\bm{k}}(\mathrm{Re}\psi_l+\mathrm{Re}\psi_G)(\phi_+^l)^2-\int_{\bm{k}}(\mathrm{Re}\psi_l+\mathrm{Re}\psi_G)(\phi_-^l)^2-2\int \psi_W\phi_+^l\phi_-^l}\nonumber\\
    &=\left(\frac{\mathrm{Re}\psi_l+\mathrm{Re}\psi_G-\psi_W}{\mathrm{Re}\psi_l+\mathrm{Re}\psi_G+\psi_W}\right)^{1/2}
\end{align}
If we expand in small coupling we get,
\begin{align}
      \gamma=1-\frac{\psi_W}{\mathrm{Re}\psi_l}+\mathcal{O}(g^4)
\end{align}

By replacing back the formulas used in \eqref{def:Wfn} we also get the expression,
\begin{align}
    \gamma-1&=\frac{g^2}{2\mathrm{Re}\psi_l}\int d\eta\sqrt{-g}\int d\eta'\sqrt{-g}K_{+}^l(k,\eta)G^s_{++}(k;\eta,\eta')K^l_{+}(k,\eta)+\mathcal{O}(g^4)\nonumber\\
    &=\frac{g^2}{2} \int d\eta\sqrt{-g}\int d\eta'\sqrt{-g} G^l_W(k;\eta,\eta')G^s_W(k;\eta,\eta')+\mathcal{O}(g^4)
\end{align}
where $G_W^{l/s} $ is the Wightman function of the system/enviroment.
\section{Massless environment}
Computing the wavefunction coefficients from \eqref{def:wfncoeffs} is generally a difficult task. To avoid technical complications and to illustrate their qualitative behavior, let us instead consider a toy model following~\cite{Colas:2022kfu,Colas:2024xjy} in which the environment is represented by a scalar field . A peculiarity of this example is that, if $\sigma$ is massless, then interactions involving fewer than two derivatives between $\sigma$ and $\varphi$ do not decay after horizon crossing. As a result, such couplings lead to infrared (IR) divergences, which in turn cause the correlation functions of $\varphi$ to grow with time.  While this procedure does not remove the secular growth, it makes explicit that the theory remains well defined, provided that perturbativity is kept under control.

\paragraph{\underline{$\alpha\dot\phi\sigma$}}
Let us first consider  the action given by
\begin{align}
S=\int d^4x\sqrt{-g}\left[-\frac{1}{2}(\partial_\mu\phi)^2-\frac{1}{2}(\partial_\mu\sigma)^2-\frac{1}{2}m^2\sigma^2-\alpha\dot\phi\sigma\right]
\end{align} 
We want to compute the density matrix after integrating out over the field $\sigma$. To do so we can use the results from the last section and replace the propagators of the enviroment by the propagators of $\sigma$ computed using the free theory\footnote{See \cite{Cespedes:2023aal} for more details on the propagators.}. For the case when $\sigma$ is massless the integrals can be done exactly. Nevertheless it will more useful to do a late time expansion.  We obtain that  $\psi^W$ and $\psi^F$ at leading order in $k/aH$ are,
\begin{align}
    \psi^W_\phi&=\frac{\alpha^2 k^3}{2H^4}\left(\frac{1}{k^2\eta_0^2}-\frac{\pi}{k\eta_0}+\log(-k\eta_0)^2+\mathcal{O}(\log(k\eta_0))\right),\\
    \psi^F_\phi&=\frac{\alpha^2 k^3}{2H^4}\left(-\frac{1}{k^2\eta_0^2}+\frac{i}{k\eta_0}\left(-4+2\gamma_E-3i\pi+2\log(-2k\eta_0)\right)+\log(-k\eta_0)^2+\mathcal{O}(\log(-k\eta_0))\right),
\end{align}
The dominant terms, which grow as a power of $k/aH$,  and as expected dominate over the tree level wavefunction coefficient (which is constant on superhorizon scales). The fact that  $\psi^W$ and $\psi^F$ grow at the same rate implies that the the growth cancel for the two point function of $\phi$, which  has only a secular growth,
\begin{align}
\langle\phi(\k)\phi(-\k)\rangle'
&=\frac{H^2}{2k^3}\left(1+\frac{\alpha^2}{H^2}\left((-2+\gamma_E+\log(-2k\eta_0))^2-\frac{\pi^2}{12}+1\right)+\mathcal{O}(k\eta_0)\right)
\end{align}
However this cancellation is not present when computing the momentum correlator which grows,
\begin{align}
    \langle\pi_\phi(\k)\pi_\phi(-\k)\rangle'=\frac{k^3}{2H^2}\frac{1}{k^2\eta_0^2}\left(\frac{1}{2}-\frac{\alpha^2}{H^2}\left(-3+\gamma_E+\log(-2k\eta_0)\right)^2+\mathcal{O}(-k\eta_0)\right)
\end{align}

The growth of the off-diagonal parts can be more clearly seen in the   Keldysh basis\footnote{In the Keldysh basis $\phi_r=1/2(\phi_1+\phi_2)$ and $\phi_a=\phi_1-\phi_2$.} where the density matrix becomes,
\begin{align}
    \rho_\phi[\phi_a,\phi_s]=\exp\left(-\frac{1}{2}\int_\k \psi_{aa}\phi_a^2-\int_\k \psi_{ar}\phi_a\phi_s-\frac{1}{2}\int_\k \psi_{ss}\phi_s^2 \right)
    \label{reducedrho_linear}
\end{align}
with
\begin{align}
    \psi_{aa}&=\frac{k^3}{4H^2}\left(1+\frac{\alpha^2}{2H^2}\frac{1}{k^2\eta_0^2}(4-12\pi k\eta_0+9\pi^2k\eta_0^2)+\mathcal{O}(-k\eta_0)\right)\\
    \psi_{as}&=\frac{i k^3}{2H^2}\frac{1}{k\eta_0}\left(1+\frac{2\alpha^2}{H^2}(-6+3\gamma_E+3\log(-2k\eta_0))+\mathcal{O}(-k\eta_0)\right)\nonumber\\
    \psi_{ss}&=\frac{k^3}{H^2}\left(1+\frac{2\alpha^2}{H^2}(-2+\gamma_E+\log(-2k\eta_0))^2+\mathcal{O}(-k\eta_0)\right).
    \label{SKcoeff_linearmixing}
\end{align}
It can be easily seen that at late times $\psi_{aa}$ grows faster than the rest of the wavefunction coefficients. 
A straightforward computation will show that the  determinant of the covariance matrix in terms of the Keldysh basis coefficients is simply,
\begin{align}
    \det \Sigma=\frac{\psi_{aa}}{\psi_{ss}}\label{detC:linearmixing}
\end{align}
which shows how the covariance matrix grows on superhorizon scales is related to the $\psi_{aa}$ coefficient. Actually we can interpret as the onset of the system becoming classical. We can see this by noticing that $\det \Sigma$ grows when the momentum correlators start to become large.

\end{appendix}

\bibliographystyle{apsrev}
\bibliography{myrefs}

\providecommand{\href}[2]{#2}\begingroup\raggedright\begin{thebibliography}{10}

\bibitem{Penrose:1980ge}
R.~Penrose, \emph{{Singularities and time asymmetry in General Relativity: an
  Einstein Centennial}}, Cambridge University Press (08, 1980),
  \href{https://doi.org/00000}{00000}.

\bibitem{Wald:2005cb}
R.M.~Wald, \emph{{The Arrow of time and the initial conditions of the
  universe}}, \href{https://doi.org/10.1016/j.shpsb.2006.03.005}{\emph{Stud.
  Hist. Phil. Mod. Phys.} {\bfseries 37} (2006) 394}
  [\href{https://arxiv.org/abs/gr-qc/0507094}{{\ttfamily gr-qc/0507094}}].

\bibitem{Carroll:2004pn}
S.M.~Carroll and J.~Chen, \emph{{Spontaneous inflation and the origin of the
  arrow of time}},  \href{https://arxiv.org/abs/hep-th/0410270}{{\ttfamily
  hep-th/0410270}}.

\bibitem{Bousso:2011aa}
R.~Bousso, \emph{{Vacuum Structure and the Arrow of Time}},
  \href{https://doi.org/10.1103/PhysRevD.86.123509}{\emph{Phys. Rev. D}
  {\bfseries 86} (2012) 123509}
  [\href{https://arxiv.org/abs/1112.3341}{{\ttfamily 1112.3341}}].

\bibitem{Wall:2010jtc}
A.C.~Wall, \emph{{The Generalized Second Law implies a Quantum Singularity
  Theorem}}, \href{https://doi.org/10.1088/0264-9381/30/19/199501}{\emph{Class.
  Quant. Grav.} {\bfseries 30} (2013) 165003}
  [\href{https://arxiv.org/abs/1010.5513}{{\ttfamily 1010.5513}}].

\bibitem{Maldacena:2012xp}
J.~Maldacena and G.L.~Pimentel, \emph{{Entanglement entropy in de Sitter
  space}}, \href{https://doi.org/10.1007/JHEP02(2013)038}{\emph{JHEP}
  {\bfseries 02} (2013) 038} [\href{https://arxiv.org/abs/1210.7244}{{\ttfamily
  1210.7244}}].

\bibitem{Brahma:2020zpk}
S.~Brahma, O.~Alaryani and R.~Brandenberger, \emph{{Entanglement entropy of
  cosmological perturbations}},
  \href{https://doi.org/10.1103/PhysRevD.102.043529}{\emph{Phys. Rev. D}
  {\bfseries 102} (2020) 043529}
  [\href{https://arxiv.org/abs/2005.09688}{{\ttfamily 2005.09688}}].

\bibitem{Boutivas:2023mfg}
K.~Boutivas, D.~Katsinis, G.~Pastras and N.~Tetradis, \emph{{Entanglement in
  cosmology}}, \href{https://doi.org/10.1088/1475-7516/2024/04/017}{\emph{JCAP}
  {\bfseries 04} (2024) 017}
  [\href{https://arxiv.org/abs/2310.17208}{{\ttfamily 2310.17208}}].

\bibitem{DuasoPueyo:2024usw}
C.~Duaso~Pueyo, H.~Goodhew, C.~McCulloch and E.~Pajer, \emph{{Perturbative
  unitarity bounds from momentum-space entanglement}},
  \href{https://doi.org/10.1007/JHEP08(2025)047}{\emph{JHEP} {\bfseries 08}
  (2025) 047} [\href{https://arxiv.org/abs/2410.23709}{{\ttfamily
  2410.23709}}].

\bibitem{Zurek:2003zz}
W.H.~Zurek, \emph{{Decoherence, einselection, and the quantum origins of the
  classical}}, \href{https://doi.org/10.1103/RevModPhys.75.715}{\emph{Rev. Mod.
  Phys.} {\bfseries 75} (2003) 715}
  [\href{https://arxiv.org/abs/quant-ph/0105127}{{\ttfamily
  quant-ph/0105127}}].

\bibitem{Calzetta:1995ys}
E.~Calzetta and B.L.~Hu, \emph{{Quantum fluctuations, decoherence of the mean
  field, and structure formation in the early universe}},
  \href{https://doi.org/10.1103/PhysRevD.52.6770}{\emph{Phys. Rev. D}
  {\bfseries 52} (1995) 6770}
  [\href{https://arxiv.org/abs/gr-qc/9505046}{{\ttfamily gr-qc/9505046}}].

\bibitem{Burgess:2006jn}
C.P.~Burgess, R.~Holman and D.~Hoover, \emph{{Decoherence of inflationary
  primordial fluctuations}},
  \href{https://doi.org/10.1103/PhysRevD.77.063534}{\emph{Phys. Rev. D}
  {\bfseries 77} (2008) 063534}
  [\href{https://arxiv.org/abs/astro-ph/0601646}{{\ttfamily
  astro-ph/0601646}}].

\bibitem{Burgess:2015ajz}
C.P.~Burgess, R.~Holman and G.~Tasinato, \emph{{Open EFTs, IR effects
  {\textbackslash}{\&} late-time resummations: systematic corrections in
  stochastic inflation}},
  \href{https://doi.org/10.1007/JHEP01(2016)153}{\emph{JHEP} {\bfseries 01}
  (2016) 153} [\href{https://arxiv.org/abs/1512.00169}{{\ttfamily
  1512.00169}}].

\bibitem{Boyanovsky:2015tba}
D.~Boyanovsky, \emph{{Effective field theory during inflation: Reduced density
  matrix and its quantum master equation}},
  \href{https://doi.org/10.1103/PhysRevD.92.023527}{\emph{Phys. Rev. D}
  {\bfseries 92} (2015) 023527}
  [\href{https://arxiv.org/abs/1506.07395}{{\ttfamily 1506.07395}}].

\bibitem{Boyanovsky:2015xoa}
D.~Boyanovsky, \emph{{Effective Field Theory out of Equilibrium: Brownian
  quantum fields}},
  \href{https://doi.org/10.1088/1367-2630/17/6/063017}{\emph{New J. Phys.}
  {\bfseries 17} (2015) 063017}
  [\href{https://arxiv.org/abs/1503.00156}{{\ttfamily 1503.00156}}].

\bibitem{Burgess:2022nwu}
C.P.~Burgess, R.~Holman, G.~Kaplanek, J.~Martin and V.~Vennin, \emph{{Minimal
  decoherence from inflation}},
  \href{https://doi.org/10.1088/1475-7516/2023/07/022}{\emph{JCAP} {\bfseries
  07} (2023) 022} [\href{https://arxiv.org/abs/2211.11046}{{\ttfamily
  2211.11046}}].

\bibitem{Colas:2022kfu}
T.~Colas, J.~Grain and V.~Vennin, \emph{{Quantum recoherence in the early
  universe}}, \href{https://doi.org/10.1209/0295-5075/acdd94}{\emph{EPL}
  {\bfseries 142} (2023) 69002}
  [\href{https://arxiv.org/abs/2212.09486}{{\ttfamily 2212.09486}}].

\bibitem{Salcedo:2024smn}
S.A.~Salcedo, T.~Colas and E.~Pajer, \emph{{The open effective field theory of
  inflation}}, \href{https://doi.org/10.1007/JHEP10(2024)248}{\emph{JHEP}
  {\bfseries 10} (2024) 248}
  [\href{https://arxiv.org/abs/2404.15416}{{\ttfamily 2404.15416}}].

\bibitem{Salcedo:2025ezu}
S.A.~Salcedo, T.~Colas, L.~Dufner and E.~Pajer, \emph{{An Open System Approach
  to Gravity}},  \href{https://arxiv.org/abs/2507.03103}{{\ttfamily
  2507.03103}}.

\bibitem{Lombardo:2005iz}
F.C.~Lombardo and D.~Lopez~Nacir, \emph{{Decoherence during inflation: The
  Generation of classical inhomogeneities}},
  \href{https://doi.org/10.1103/PhysRevD.72.063506}{\emph{Phys. Rev. D}
  {\bfseries 72} (2005) 063506}
  [\href{https://arxiv.org/abs/gr-qc/0506051}{{\ttfamily gr-qc/0506051}}].

\bibitem{Campo:2008ju}
D.~Campo and R.~Parentani, \emph{{Decoherence and entropy of primordial
  fluctuations. I: Formalism and interpretation}},
  \href{https://doi.org/10.1103/PhysRevD.78.065044}{\emph{Phys. Rev. D}
  {\bfseries 78} (2008) 065044}
  [\href{https://arxiv.org/abs/0805.0548}{{\ttfamily 0805.0548}}].

\bibitem{Campo:2008ij}
D.~Campo and R.~Parentani, \emph{{Decoherence and entropy of primordial
  fluctuations II. The entropy budget}},
  \href{https://doi.org/10.1103/PhysRevD.78.065045}{\emph{Phys. Rev. D}
  {\bfseries 78} (2008) 065045}
  [\href{https://arxiv.org/abs/0805.0424}{{\ttfamily 0805.0424}}].

\bibitem{Burgess:2014eoa}
C.P.~Burgess, R.~Holman, G.~Tasinato and M.~Williams, \emph{{EFT Beyond the
  Horizon: Stochastic Inflation and How Primordial Quantum Fluctuations Go
  Classical}}, \href{https://doi.org/10.1007/JHEP03(2015)090}{\emph{JHEP}
  {\bfseries 03} (2015) 090} [\href{https://arxiv.org/abs/1408.5002}{{\ttfamily
  1408.5002}}].

\bibitem{Burgess:2020tbq}
C.P.~Burgess, \emph{{Introduction to Effective Field Theory}}, Cambridge
  University Press (12, 2020),
  \href{https://doi.org/10.1017/9781139048040}{10.1017/9781139048040}.

\bibitem{Colas:2024xjy}
T.~Colas, C.~de~Rham and G.~Kaplanek, \emph{{Decoherence out of fire: purity
  loss in expanding and contracting universes}},
  \href{https://doi.org/10.1088/1475-7516/2024/05/025}{\emph{JCAP} {\bfseries
  05} (2024) 025} [\href{https://arxiv.org/abs/2401.02832}{{\ttfamily
  2401.02832}}].

\bibitem{PhysRevA.72.052113}
F.M.~Cucchietti, J.P.~Paz and W.H.~Zurek, \emph{Decoherence from spin
  environments}, \href{https://doi.org/10.1103/PhysRevA.72.052113}{\emph{Phys.
  Rev. A} {\bfseries 72} (2005) 052113}.

\bibitem{Gorbenko:2019rza}
V.~Gorbenko and L.~Senatore, \emph{{$\lambda \phi^4$ in dS}},
  \href{https://arxiv.org/abs/1911.00022}{{\ttfamily 1911.00022}}.

\bibitem{Cespedes:2023aal}
S.~C\'espedes, A.-C.~Davis and D.-G.~Wang, \emph{{On the IR divergences in de
  Sitter space: loops, resummation and the semi-classical wavefunction}},
  \href{https://doi.org/10.1007/JHEP04(2024)004}{\emph{JHEP} {\bfseries 04}
  (2024) 004} [\href{https://arxiv.org/abs/2311.17990}{{\ttfamily
  2311.17990}}].

\bibitem{Cohen:2021fzf}
T.~Cohen, D.~Green, A.~Premkumar and A.~Ridgway, \emph{{Stochastic Inflation at
  NNLO}}, \href{https://doi.org/10.1007/JHEP09(2021)159}{\emph{JHEP} {\bfseries
  09} (2021) 159} [\href{https://arxiv.org/abs/2106.09728}{{\ttfamily
  2106.09728}}].

\bibitem{Kamenev_2011}
A.~Kamenev, \emph{Field Theory of Non-Equilibrium Systems}, Cambridge
  University Press (2011).

\bibitem{Serafini:2003ke}
A.~Serafini, F.~Illuminati and S.~De~Siena, \emph{{Von Neumann entropy, mutual
  information and total correlations of Gaussian states}},
  \href{https://doi.org/10.1088/0953-4075/37/2/L02}{\emph{J. Phys. B}
  {\bfseries 37} (2004) L21}
  [\href{https://arxiv.org/abs/quant-ph/0307073}{{\ttfamily
  quant-ph/0307073}}].

\bibitem{Lopez:2025arw}
F.~Lopez and N.~Bartolo, \emph{{Quantum signatures and decoherence during
  inflation from deep subhorizon perturbations}},
  \href{https://arxiv.org/abs/2503.23150}{{\ttfamily 2503.23150}}.

\bibitem{Burgess:2025dwm}
C.P.~Burgess, R.~Holman and G.~Kaplanek, \emph{{Inflationary Decoherence from
  the Gravitational Floor}},
  \href{https://arxiv.org/abs/2509.07769}{{\ttfamily 2509.07769}}.

\bibitem{Martin:2018zbe}
J.~Martin and V.~Vennin, \emph{{Observational constraints on quantum
  decoherence during inflation}},
  \href{https://doi.org/10.1088/1475-7516/2018/05/063}{\emph{JCAP} {\bfseries
  05} (2018) 063} [\href{https://arxiv.org/abs/1801.09949}{{\ttfamily
  1801.09949}}].

\bibitem{Weinberg:qm}
S.~Weinberg, \emph{{Lectures on Quantum Mechanics, 2nd Edition}}, Cambridge
  University Press (11, 2015).

\bibitem{CALABRESE201831}
P.~Calabrese, \emph{Entanglement and thermodynamics in non-equilibrium isolated
  quantum systems},
  \href{https://doi.org/https://doi.org/10.1016/j.physa.2017.10.011}{\emph{Physica
  A: Statistical Mechanics and its Applications} {\bfseries 504} (2018) 31}.

\bibitem{Kofman:1997yn}
L.~Kofman, A.D.~Linde and A.A.~Starobinsky, \emph{{Towards the theory of
  reheating after inflation}},
  \href{https://doi.org/10.1103/PhysRevD.56.3258}{\emph{Phys. Rev. D}
  {\bfseries 56} (1997) 3258}
  [\href{https://arxiv.org/abs/hep-ph/9704452}{{\ttfamily hep-ph/9704452}}].

\bibitem{Bousso:2006ge}
R.~Bousso, B.~Freivogel and I.-S.~Yang, \emph{{Eternal Inflation: The Inside
  Story}}, \href{https://doi.org/10.1103/PhysRevD.74.103516}{\emph{Phys. Rev.
  D} {\bfseries 74} (2006) 103516}
  [\href{https://arxiv.org/abs/hep-th/0606114}{{\ttfamily hep-th/0606114}}].

\bibitem{Arkani-Hamed:2007ryv}
N.~Arkani-Hamed, S.~Dubovsky, A.~Nicolis, E.~Trincherini and G.~Villadoro,
  \emph{{A Measure of de Sitter entropy and eternal inflation}},
  \href{https://doi.org/10.1088/1126-6708/2007/05/055}{\emph{JHEP} {\bfseries
  05} (2007) 055} [\href{https://arxiv.org/abs/0704.1814}{{\ttfamily
  0704.1814}}].

\bibitem{DAlessio:2015qtq}
L.~D'Alessio, Y.~Kafri, A.~Polkovnikov and M.~Rigol, \emph{{From quantum chaos
  and eigenstate thermalization to statistical mechanics and thermodynamics}},
  \href{https://doi.org/10.1080/00018732.2016.1198134}{\emph{Adv. Phys.}
  {\bfseries 65} (2016) 239}
  [\href{https://arxiv.org/abs/1509.06411}{{\ttfamily 1509.06411}}].

\bibitem{Berges:2004yj}
J.~Berges, \emph{{Introduction to nonequilibrium quantum field theory}},
  \href{https://doi.org/10.1063/1.1843591}{\emph{AIP Conf. Proc.} {\bfseries
  739} (2004) 3} [\href{https://arxiv.org/abs/hep-ph/0409233}{{\ttfamily
  hep-ph/0409233}}].

\bibitem{Vilenkin:1982de}
A.~Vilenkin, \emph{{Creation of Universes from Nothing}},
  \href{https://doi.org/10.1016/0370-2693(82)90866-8}{\emph{Phys. Lett. B}
  {\bfseries 117} (1982) 25}.

\bibitem{Hartle:1983ai}
J.B.~Hartle and S.W.~Hawking, \emph{{Wave Function of the Universe}},
  \href{https://doi.org/10.1103/PhysRevD.28.2960}{\emph{Phys. Rev. D}
  {\bfseries 28} (1983) 2960}.

\bibitem{Ivo:2024ill}
V.~Ivo, Y.-Z.~Li and J.~Maldacena, \emph{{The no boundary density matrix}},
  \href{https://doi.org/10.1007/JHEP02(2025)124}{\emph{JHEP} {\bfseries 02}
  (2025) 124} [\href{https://arxiv.org/abs/2409.14218}{{\ttfamily
  2409.14218}}].

\bibitem{Balasubramanian:2011wt}
V.~Balasubramanian, M.B.~McDermott and M.~Van~Raamsdonk, \emph{{Momentum-space
  entanglement and renormalization in quantum field theory}},
  \href{https://doi.org/10.1103/PhysRevD.86.045014}{\emph{Phys. Rev. D}
  {\bfseries 86} (2012) 045014}
  [\href{https://arxiv.org/abs/1108.3568}{{\ttfamily 1108.3568}}].

\end{thebibliography}\endgroup

\end{document}